\documentclass[%
 aip,
 amsmath,amssymb,
preprint,%
]{revtex4-1}

\usepackage{graphicx}
\usepackage{dcolumn}
\usepackage{bm}

\usepackage[utf8]{inputenc}
\usepackage[T1]{fontenc}
\usepackage{mathptmx}
\usepackage{etoolbox}
\usepackage{makecell}
\usepackage{colortbl}

\newcommand{\bea}{\begin{eqnarray}}
\newcommand{\eea}{\end{eqnarray}}

\makeatletter
\def\@email#1#2{%
 \endgroup
 \patchcmd{\titleblock@produce}
  {\frontmatter@RRAPformat}
  {\frontmatter@RRAPformat{\produce@RRAP{*#1\href{mailto:#2}{#2}}}\frontmatter@RRAPformat}
  {}{}
}%
\makeatother
\begin{document}


\title[]{Three Types of Non-Fermi-Liquid Fixed Point\\for a Triplet Quantum Impurity in a Cubic Metal}
\author{Anna I.\ T\'oth}
 \email{atoth2@ed.ac.uk}
\affiliation{School of Physics and Astronomy, Centre for Science at Extreme Conditions,\\ The University of Edinburgh,\\ Edinburgh EH9 3FD, United Kingdom
}%

\date{\today}

\begin{abstract}
  In cubic metals, a local, magnetic moment with a triplet ground state 
  coupled to $\Gamma_8$ conduction electrons can give rise to various non-Fermi liquid (NFL) quantum critical behaviors.  To date, only those exchange couplings have been studied that are spherically symmetric already in the high-temperature,
  local moment regime.  Namely, only the effects of potential scattering, dipolar spin exchange, i.e.\ Kondo, and  quadrupolar exchange couplings were considered,
  and two types of NFL fixed points have been identified.  However, in cubic symmetry, six independent exchange couplings can be present in the Hamiltonian: in addition to the spherically symmetric potential scattering and Kondo exchange,  there are four independent, spherical-symmetry-breaking Hamiltonian terms: two quadrupolar-quadrupolar, a dipolar-octupolar and a quadrupolar-octupolar exchange interaction, where the first/second element of the compound adjectives refers to the impurity/conduction-electron transitions.  While all of them flow to fixed points where rotational invariance is recovered, 
  I found that one of the quadrupolar-quadrupolar couplings flows to a previously unidentified NFL fixed point.
  I derive the exchange couplings allowed by cubic symmetry, 
  solve them with the numerical renormalization group, and 
  present three types of NFL excitation spectrum\textemdash one of which is novel, at least in the context of a quantum impurity with a triplet ground state.


\end{abstract}

\maketitle

\section{\label{sec:level1}Introduction}

The study of the resistivity of metals at low temperatures 
has led to groundbreaking discoveries, including superconductivity and the integer and fractional quantum Hall effects. In 1934, physicists found an anomalous increase in the resistance of gold as the temperature was lowered \cite{deHaas34}. The phenomenon, known as the Kondo
effect, comes from the interaction between conduction electrons and individual magnetic moments. It
is described by a so-called quantum impurity or Kondo model \cite{Kondo64}. 
In many cases\textemdash sometimes even in the presence of magnetic impurities\textemdash the standard theory of electrons in
metals at low temperatures, Landau’s Fermi liquid theory \cite{Landau56}, very successfully accounts for their basic
properties, as e.g.\ the quadratic temperature-dependence of the electrical resistance.
Its greatest success was its ability to describe many
heavy fermion metals, whose name derives from the huge apparent masses acquired by their conduction
electrons \cite{Fisk91}. 
However, increasingly many metals have been synthesized where Fermi-liquid predictions fail. 
Metals that do not conform to Fermi liquid theory are called exotic/strange/bad or simply non-Fermi liquids (NFLs).
NFL phenomena observed in three- and low-dimensional materials seem abundant and diverse,
 yet only a few NFL models are known that can be solved analytically or numerically.

NFL physics in certain heavy fermion compounds is manifest e.g.\ 
as a divergence, logarithmic or mild power-law $T$-dependence of the low-temperature 
electronic specific heat coefficient and the magnetic susceptibility, as well as in the subquadratic temperature dependence of the resistivity \cite{Stewart01}.
In some heavy fermion systems, this type of  NFL behavior has been attributed to a single-site,
multichannel Kondo effect due to quantum impurities \cite{Nozieres80}. Most notably, Cox proposed a time-reversal-symmetry-breaking two-channel Kondo (2CK) effect arising in uranium-based, cubic 
heavy fermion systems, where the U atoms have an $f^2$-electronic ground state configuration with a $\Gamma_3$ non-Kramers doublet ground state that hybridizes 
with local, $\Gamma_8$ conduction electrons \cite{Cox87,Toth25}. The experimental realization of the 2CK effect in cubic (and even in tetragonal) 4 and $5f$-electron systems has been pursued
by many groups over many years, and in 2018,
Y$_{1-x}$Pr$_x$Ir$_2$Zn$_{20}$ \cite{Yamane18,Yanagisawa19} emerged as the most widely accepted realization of 2CK physics in a cubic, heavy fermion compound\textemdash decades after Cox's original proposal \cite{Cox87}. In Ref.\ \onlinecite{Toth25}, we listed every type of NFL behavior that can occur in cubic metals due to doublet impurities without any accidental degeneracy of the conduction electron levels and their hybridization with the local moment, and argued that, in the absence of accidental degeneracy, only systems with higher point group symmetries such as e.g.\ the cubic point group, $O$ (using the Sch\"onflies symbols for point groups), that have at least one three-dimensional irreducible representation (irrep) \cite{Koster63}, can host NFL quantum impurity physics.
Apart from the 2CK behavior derived by Cox, we found that two more types of NFL quantum critical behavior can exist in diluted, cubic compounds without accidental degeneracy, namely, the topological Kondo (spin-half impurity, spin-one conduction electron) \cite{Fabrizio94,Beri12} and the spin-half impurity, spin-$\frac 3 2$ conduction electron Kondo effects.

In Refs.\ \onlinecite{Fabrizio96,Sengupta96}, a mapping was established between higher-spin-conduction-electron multichannel Kondo models and the original, spin-half conduction electron multichannel Kondo model introduced in Ref.\ \onlinecite{Nozieres80}.
According to this mapping, the topological Kondo model yields identical results to the four-channel spin-half conduction electron Kondo model\textemdash e.g.\ for the impurity contribution to the specific heat and the magnetic susceptibility\textemdash whereas one channel of spin-$\frac 3 2$ conduction electrons can be mapped onto ten channels of spin-half conduction electrons. Thus, all three NFL quantum impurity behaviors that are cubic-symmetry-protected can be matched with different overscreened, multichannel Kondo models with spin-half conduction electrons. This finding raises the question of whether there exist any NFL quantum impurity models---symmetry-protected or otherwise---that are not of multichannel Kondo type. This question has been answered in the affirmative in Ref.\ \onlinecite{Patri20}, where a cubic, NFL, exchange Hamiltonian was constructed by overscreening a $\Gamma_3$, non-Kramers doublet impurity with $\Gamma_6$ and $\Gamma_8$ conduction electrons simultaneously. However, the emergence of the corresponding NFL quantum impurity effect in real materials is not likely to occur as it would require the accidental degeneracy between the $\Gamma_6$ and $\Gamma_8$ screening channels that are not equivalent by symmetry \cite{Toth11,Toth25}.  Nevertheless, the corresponding NFL behavior has not been matched with a multichannel Kondo model yet.  

In this paper, I approach the same question but applied to a spin-one impurity by examining what types of NFL can arise in cubic metals. For triplet impurities, so far only those exchange couplings were studied that are spherically symmetric already in the high-temperature, local moment regime \cite{Koga99,Koga19}. Namely, so far only the effects of
potential scattering, Kondo exchange and quadrupolar exchange couplings were considered
and correspondingly two types of NFL behavior were identified. Here I extend these studies by deriving also those exchange couplings that only have cubic symmetry
in the high-temperature, local moment regime and solving them with the numerical renormalization group \cite{Wilson75} (NRG) with the aim of identifying all types of quantum critical behavior that might appear in  (cubic) metals due to single impurities.

As the 32 point groups and the corresponding double groups only have four-dimensional irreps at most \cite{Koster63}, it means that to achieve overscreening without accidental degeneracy, the impurity state can only be two- or at most threefold degenerate, as the degeneracy of the local conduction electron states needs to exceed that of the impurity. In Ref.\ \onlinecite{Toth25}, we analyzed the case of twofold impurity degeneracy. Here I complete the analysis by studying the case of threefold degeneracy.

As shown by the subsequent analysis, for a triplet impurity, there are altogether six independent exchange couplings allowed by cubic symmetry,
and while each of them flows to a quantum impurity fixed point where rotational invariance is recovered,
I found that one coupling flows to a so far unidentified NFL fixed point.  This represents the third distinct NFL universality class available for
  triplet quantum impurities in a cubic environment, complementing the two known cases. The importance of this result lies in its broadening of the theoretical landscape of impurity-driven NFL behavior in strongly correlated electron systems, suggesting that cubic environments may host more diverse quantum
  critical phenomena than captured by the conventional, overscreened, multichannel Kondo models.

The paper is organized as follows. In Sec.\ \ref{sec:constr}, I construct all, cubic-symmetry-allowed exchange couplings, and to provide a more complete picture, discuss how they transform under time-reversal symmetry.  In Subsec.\ \ref{subsec:SpherSym}, 
those exchange couplings are constructed that, in addition to having cubic symmetry, are also spherical invariants. These couplings have already been identified and studied in Refs.\ \onlinecite{Koga99,Koga19}. I solve these models using NRG and present their NFL, finite-size spectrum.
Then in Subsec.\ \ref{subsec:CubSym}, all exchange couplings with cubic symmetry are constructed, compared to their spherically symmetric counterparts and solved with NRG. A so far unidentified NFL fixed point in the context of triplet impurities in metals is found. Finally, conclusions are drawn in Sec.\ \ref{sec:end}.  

\section{\label{sec:constr}Construction}

For a triplet impurity ground state and without any accidental degeneracy of the local conduction electron couplings and levels, at least fourfold degenerate local conduction electron states are required to construct  Kondo-type exchange Hamiltonians with built-in frustration that\textemdash together with the kinetic energy of the conduction electrons\textemdash lead to NFL physics.  The only point groups with both three- and four-dimensional irreps are $O,T_d,O_h$ and the corresponding double groups \cite{Koster63}. I concentrate on the octahedral or cubic point group $O$ (which is isomorphic to $T_d$) as it yields every possible cubic exchange Hamiltonian due to its identical Clebsch--Gordan coefficients with the other symmetries \cite{Koster63}.
 The impurity is assumed to have an even-electron, $\Gamma_4$ (i.e.\ $T_1$, using a Mulliken symbol) ground state. This choice does not result in any loss of generality as the only other three-dimensional irrep of $O$ is $\Gamma_5$, and the cubic Clebsch--Gordan coefficients used for the Hamiltonian construction coming from either  $\Gamma_4\otimes\Gamma_4$ or $\Gamma_5\otimes\Gamma_5$  are the same.  The local conduction electrons are assumed to transform as a Kramers quartet, i.e.\ the $\Gamma_8$ irrep of $O^\prime$, the cubic double group. 

The construction of all independent Kondo-type exchange Hamiltonian terms proceeds by creating every possible irreducible, electron-hole tensor operator from the local conduction electrons, as well as creating every possible irreducible tensor operator from the impurity ket and bra states. Every scalar product of the impurity and local conduction electron-hole irreducible tensor operators that transform in the same way under cubic rotations yields an independent Hamiltonian term. 

\subsection{\label{subsec:SpherSym} Exchange couplings with spherical symmetry}

In this subsection, I construct those Kondo-type exchange Hamiltonians that are not only cubic but also spherical invariants. Note that in a cubic field, neither a $J=1$ nor a $J=\frac 3 2$ multiplet breaks. They correspond to $\Gamma_4$ and $\Gamma_8$ irreps, respectively. 
I introduce the following two, ket and bra multiplets that are both eigenstates of the operators $\left({\vec S}^{\textrm{\,imp}}\right)^{\,2},\,S^{\textrm{\,imp}}_z$,  
\bea\label{eq:spin1multiplet}
\left(
\begin{array}{l}
  |1\quad\,1\rangle\\
  |1\quad\,0\rangle\\
  |1\,-1\rangle
\end{array}
\right)\,,
\quad
\left(
\begin{array}{r}
  \langle1\,-1|\\
  -\langle1\quad\,0|\\
  \langle1\quad\,1|
\end{array}
\right)\,.
\eea
Within each $S=1$ multiplet, the same row comprises eigenvectors with the same eigenvalue, and neighboring rows are connected in the same way by the ladder operators.

With the choice
      \bea
      {\mathcal T}\,=\,-\,i\sigma_y\,\otimes\,{\mathcal K}
      \eea
      for the time-reversal operator, ${\mathcal T}$,  where $i\sigma_y$ (with $\sigma_y$ the Pauli matrix) acts on the spin-half part (assuming that higher, half-integer spin objects are constructed from spin-half electrons) and ${\mathcal K}$ denotes the complex conjugation, these states transform under time-reversal either as
      \bea
      {\mathcal T}  |1\,\,\,\, S_z\rangle = (-)^{\,S_z}\, |1\,\,\,-S_z\rangle\,,
      \eea
      if they are pure spherical harmonics, or as
            \bea
      {\mathcal T}  |1\,\,\,\,S_z\rangle = (-)^{\,S_z\,+\,1}\, |1\,\,\,-S_z\rangle\,,
      \eea
      if they come e.g.\ from the tensor product of two half-integer spin multiplets.

Based on their spherical tensor product decomposition
\bea\label{eq:1times1}
(S=1)\otimes 1=0\oplus 1\oplus 2\,,
\eea
the following rank-0 (scalar), 1 (vector) and 2 irreducible tensor operators, i.e.\ the impurity number, spin and quadrupole operators, can be formed
\bea\label{eq:impno}
n^{\textrm{\,imp}}&\equiv&|1\quad\,1\rangle \langle1\quad\,1|\,+\,|1\quad\,0\rangle \langle1\quad\,0|\,+\,|1\,-1\rangle\langle1\,-1|\\
\left(\begin{array}{c}
  S^{\textrm{\,imp}}_1\\
  S^{\textrm{\,imp}}_0\\
  S^{\textrm{\,imp}}_{-1}
\end{array}\right)&\equiv&\left(\begin{array}{c}
  -\,S^{\textrm{\,imp}}_+/\sqrt{2}\\
  S^{\textrm{\,imp}}_z\\
  S^{\textrm{\,imp}}_{-}/\sqrt{2}
\end{array}\right)\,\equiv\,\left(\begin{array}{r}
  -|1\quad\,1\rangle \langle1\quad\,0|\,-\,|1\quad\,0\rangle \langle1\,-1|\\
  |1\quad\,1\rangle \langle1\quad\,1|\,-\,|1\,-1\rangle \langle1\,-1|\\
  |1\,-1\rangle \langle1\quad\,0|\,+\,|1\quad\,0\rangle \langle1\quad\,1|
\end{array}\right)
\eea
\begin{multline}\label{eq:QP_imp}
\left(\begin{array}{c}
  Q^{\textrm{\,imp}}_2\\
  Q^{\textrm{\,imp}}_1\\
  Q^{\textrm{\,imp}}_{0}\\
  Q^{\textrm{\,imp}}_{-1}\\
  Q^{\textrm{\,imp}}_{-2}
\end{array}\right)\equiv
\left(\begin{array}{c}
  |1\quad\,1\rangle \langle1\,-1|\\
  -\frac{1}{\sqrt 2}\left(|1\quad\,1\rangle \langle1\quad\,0|\,-\,|1\quad\,0\rangle \langle1\,-1|\right)\\
  \frac{1}{\sqrt 6}\left(|1\quad\,1\rangle \langle1\quad\,1|\,-\,2\,|1\quad\,0\rangle \langle1\quad\,0|\,+\,|1\,-1\rangle \langle1\,-1|\right)\\
  \frac{1}{\sqrt 2}\left(|1\quad\,0\rangle \langle1\quad\,1|\,-\,|1\,-1\rangle \langle1\quad\,0|\right)\\
  |1\,-1\rangle \langle1\quad\,1|
\end{array}\right)\\=
\frac 1 2\left[\begin{array}{c}
 \left(S^{\textrm{\,imp}}_x\right)^2\,-\,\left(S^{\textrm{\,imp}}_y\right)^2\,+\,i\,\overline{S^{\textrm{\,imp}}_x\,S^{\textrm{\,imp}}_y}\\
 -\left(\overline{S^{\textrm{\,imp}}_xS^{\textrm{\,imp}}_z\,+\,i\,S^{\textrm{\,imp}}_yS^{\textrm{\,imp}}_z}\right)\\
\sqrt{\frac{2}{3}}\,\left(2\,\left(S^{\textrm{\,imp}}_z\right)^2\,-\,\left(S^{\textrm{\,imp}}_x\right)^2\,-\,\left(S^{\textrm{\,imp}}_y\right)^2 \right)\\
 \overline{S^{\textrm{\,imp}}_xS^{\textrm{\,imp}}_z\,-\,i\,S^{\textrm{\,imp}}_yS^{\textrm{\,imp}}_z}\\
\left(S^{\textrm{\,imp}}_x\right)^2\,-\,\left(S^{\textrm{\,imp}}_y\right)^2\,-\,i\,\overline{S^{\textrm{\,imp}}_x\,S^{\textrm{\,imp}}_y}
\end{array}\right]
\,.
\end{multline}
    The number operator is time-reversal invariant. With the canonical definitions
\bea
      S^{\textrm{\,imp}}_x\,\equiv\,\left(S^{\textrm{\,imp}}_+\,+\,S^{\textrm{\,imp}}_-\right)/2\,,\quad S^{\textrm{\,imp}}_y\,\equiv\,\left(S^{\textrm{\,imp}}_+\,-\,S^{\textrm{\,imp}}_-\right)/{2i}\,,\quad\textrm{and}\quad{\vec S}^{\textrm{\,imp}}\equiv\left(\begin{array}{c}
        S^{\textrm{\,imp}}_x\\
        S^{\textrm{\,imp}}_y\\
        S^{\textrm{\,imp}}_z
      \end{array}\right)\,,
      \eea
      the $x,y,z$ components of the spin operator change sign under time-reversal, whereas the components of the quadrupole operator transform as
      $\,{\mathcal T}\,Q^{\textrm{\,imp}}_{k}\,{\mathcal T}^{-1}\,=\,(-)^{k}\,Q^{\textrm{\,imp}}_{-k}\,$. As a corollary of the Wigner--Eckart theorem, the quadrupole operator can also be expressed in terms of the impurity spin operator, as on the right-hand side of Eq.\ \eqref{eq:QP_imp}, where a bar over products of two $S^{\textrm{\,imp}}_k$s represents a symmetrized product with respect to different indices.

The impurity is assumed to be screened by local $\Gamma_8$ conduction electrons that transform as $J=\frac 3 2$ quartets under continuous rotations. These conduction electrons are created/annihilated  by the following irreducible tensor operator quartets
\bea\label{eq:conde_quartet}
{\bm \Psi^{\Gamma_8\,\dagger}}\equiv\left(
\renewcommand{\arraystretch}{1.35}
\begin{array}{c}
\psi_{\frac 3 2}^{\,\Gamma_8\,\dagger}\\
\psi_{\frac 1 2}^{\,\Gamma_8\,\dagger}\\
\psi_{-\frac 1 2}^{\,\Gamma_8\,\dagger}\\
\psi_{-\frac 3 2}^{\,\Gamma_8\,\dagger}
\end{array}
\renewcommand{\arraystretch}{1}
\right)\,,\quad\quad\quad\quad
      {\bm \Psi_{}^{\Gamma_8}}\equiv\left(
      \renewcommand{\arraystretch}{1.35}
\begin{array}{c}
{\psi_{-\frac 3 2}^{\,\Gamma_8\,}}\\
-{\psi_{-\frac 1 2}^{\,\Gamma_8\,}}\\
{\psi_{\frac 1 2}^{\,\Gamma_8\,}}\\
-{\psi_{\frac 3 2}^{\,\Gamma_8\,}}
\end{array}
\renewcommand{\arraystretch}{1}
\right)\,,
\eea
that transform as spinor irreps, and where the components are defined by $\psi_{J_z}^{\,\Gamma_8\,\dagger}\equiv\sum_{\vec k}\,c_{\,J=\frac 3 2\,J_z}^{\,\dagger}({\vec k})$ with $c_{\,J=\frac 3 2\,J_z}^{\,\dagger}({\vec k})$ creating a conduction electron with quantum numbers $J=\frac 3 2, J_z$ and wave vector ${\vec k}$.

The quartet components are connected by time reversal either as
$\,{\cal T}\,\Psi_{\,\left|J_z\right|}^{\Gamma_8\,\,\dagger}\,{\cal T}^{-1}=\,(-)^{\,J_z-\frac 12\,}\,\Psi_{\,-\left|J_z\right|}^{\Gamma_8\,\dagger}\,$
or as $\,{\cal T}\,\Psi_{\,\left|J_z\right|}^{\Gamma_8\,\,\dagger}\,{\cal T}^{-1}=\,(-)^{\,J_z+\frac 12\,}\,\Psi_{\,-\left|J_z\right|}^{\Gamma_8\,\dagger}\,$,
depending on the origin of the quartet.

The tensor product of the single-electron creation and annihilation irreducible tensor operators can be decomposed as
\bea
(J=\frac 3 2)\otimes \frac 3 2=0\oplus 1\oplus 2 \oplus 3\,.
\eea
Comparing this with Eq.\ \eqref{eq:1times1} gives three independent exchange couplings that can appear in a
spherically symmetric Hamiltonian, as only the scalar products of identical irreps create scalars. That is, to create spherically symmetric exchange couplings only the rank-0, 1 and 2 irreducible, electron-hole tensor operators are needed to be constructed, that correspond to the conduction electron number,
spin and quadrupole operators 
\bea\label{eq:condeno}
n^{\textrm{\,c}}&\equiv&\psi_{\frac 3 2}^{\,\Gamma_8\,\dagger\,}\psi_{\frac 3 2}^{\,\Gamma_8}\,+\,\psi_{\frac 1 2}^{\,\Gamma_8\,\dagger\,}\psi_{\frac 1 2}^{\,\Gamma_8}\,+\,\psi_{-\frac 1 2}^{\,\Gamma_8\,\dagger\,}\psi_{-\frac 1 2}^{\,\Gamma_8}\,+\,\psi_{-\frac 3 2}^{\,\Gamma_8\,\dagger\,}\psi_{-\frac 3 2}^{\,\Gamma_8}
\eea
\begin{multline}
\left(\begin{array}{c}
  J^{\textrm{\,c}}_1\\
  J^{\textrm{\,c}}_0\\
  J^{\textrm{\,c}}_{-1}
\end{array}\right)\,\equiv\,\left(\begin{array}{c}
  -J^{\textrm{\,c}}_+/\sqrt{2}\\
  J^{\textrm{\,c}}_z\\
  J^{\textrm{\,c}}_{-}/\sqrt{2}
\end{array}\right)
\equiv\left(\begin{array}{c}
  -\sqrt{\frac 3 2}\,\psi_{\frac 3 2}^{\,\Gamma_8\,\dagger\,}\psi_{\frac 1 2}^{\,\Gamma_8}\,-\,\sqrt{2}\,\psi_{\frac 1 2}^{\,\Gamma_8\,\dagger\,}\psi_{-\frac 1 2}^{\,\Gamma_8}\,-\,\sqrt{\frac 3 2}\,\psi_{-\frac 1 2}^{\,\Gamma_8\,\dagger\,}\psi_{-\frac 3 2}^{\,\Gamma_8}\\
  \frac 3 2\, \psi_{\frac 3 2}^{\,\Gamma_8\,\dagger\,}\psi_{\frac 3 2}^{\,\Gamma_8}\,+\,\frac 1 2\, \psi_{\frac 1 2}^{\,\Gamma_8\,\dagger\,}\psi_{\frac 1 2}^{\,\Gamma_8}\,-\,\frac 1 2\, \psi_{-\frac 1 2}^{\,\Gamma_8\,\dagger\,}\psi_{-\frac 1 2}^{\,\Gamma_8}\,-\,\frac 3 2\, \psi_{-\frac 3 2}^{\,\Gamma_8\,\dagger\,}\psi_{-\frac 3 2}^{\,\Gamma_8}\\
  \sqrt{\frac 3 2}\,\psi_{-\frac 3 2}^{\,\Gamma_8\,\dagger\,}\psi_{-\frac 1 2}^{\,\Gamma_8}\,+\,\sqrt{2}\,\psi_{-\frac 1 2}^{\,\Gamma_8\,\dagger\,}\psi_{\frac 1 2}^{\,\Gamma_8}\,+\,\sqrt{\frac 3 2}\,\psi_{\frac 1 2}^{\,\Gamma_8\,\dagger\,}\psi_{\frac 3 2}^{\,\Gamma_8}
\end{array}\right)
\end{multline}
\begin{multline}\label{eq:Q}
\left(\begin{array}{c}
  Q^{\textrm{\,c}}_2\\
  Q^{\textrm{\,c}}_1\\
  Q^{\textrm{\,c}}_{0}\\
  Q^{\textrm{\,c}}_{-1}\\
  Q^{\textrm{\,c}}_{-2}
\end{array}\right)\equiv-\frac{1}{\sqrt{2}}
\left[\begin{array}{c}
  \psi_{\frac 3 2}^{\,\Gamma_8\,\dagger\,}\psi_{-\frac 1 2}^{\,\Gamma_8}\,+\,\psi_{\frac 1 2}^{\,\Gamma_8\,\dagger\,}\psi_{-\frac 3 2}^{\,\Gamma_8}\\
  -\,\psi_{\frac 3 2}^{\,\Gamma_8\,\dagger\,}\psi_{\frac 1 2}^{\,\Gamma_8}\,+\,\psi_{-\frac 1 2}^{\,\Gamma_8\,\dagger\,}\psi_{-\frac 3 2}^{\,\Gamma_8}\\
  \frac{1}{\sqrt 2}\left(\psi_{\frac 3 2}^{\,\Gamma_8\,\dagger\,}\psi_{\frac 3 2}^{\,\Gamma_8}\,-\,\psi_{\frac 1 2}^{\,\Gamma_8\,\dagger\,}\psi_{\frac 1 2}^{\,\Gamma_8}\,-\,\psi_{-\frac 1 2}^{\,\Gamma_8\,\dagger\,}\psi_{-\frac 1 2}^{\,\Gamma_8}\,+\,\psi_{-\frac 3 2}^{\,\Gamma_8\,\dagger\,}\psi_{-\frac 3 2}^{\,\Gamma_8}\right)\\
   \psi_{\frac 1 2}^{\,\Gamma_8\,\dagger\,}\psi_{\frac 3 2}^{\,\Gamma_8}\,-\,\psi_{-\frac 3 2}^{\,\Gamma_8\,\dagger\,}\psi_{-\frac 1 2}^{\,\Gamma_8}\\
  \psi_{-\frac 1 2}^{\,\Gamma_8\,\dagger\,}\psi_{\frac 3 2}^{\,\Gamma_8}\,+\,\psi_{-\frac 3 2}^{\,\Gamma_8\,\dagger\,}\psi_{\frac 1 2}^{\,\Gamma_8}
  \end{array}\right]\\
=\frac 1 2\left[\begin{array}{c}
 \left(J^{\textrm{\,c}}_x\right)^2\,-\,\left(J^{\textrm{\,c}}_y\right)^2\,+\,i\,\overline{J^{\textrm{\,c}}_x\,J^{\textrm{\,c}}_y}\\
-\, \overline{J^{\textrm{\,c}}_+\,J^{\textrm{\,c}}_z}\\
\sqrt{\frac{2}{3}}\,\left(2\,\left(J^{\textrm{\,c}}_z\right)^2\,-\,\left(J^{\textrm{\,c}}_x\right)^2\,-\,\left(J^{\textrm{\,c}}_y\right)^2 \right)\\
 \overline{J^{\textrm{\,c}}_-\,J^{\textrm{\,c}}_z}\\
\left(J^{\textrm{\,c}}_x\right)^2\,-\,\left(J^{\textrm{\,c}}_y\right)^2\,-\,i\,\overline{J^{\textrm{\,c}}_x\,J^{\textrm{\,c}}_y}
\end{array}\right].
\end{multline}
The conduction electron number operator is time-reversal even, whereas $\,{\mathcal T}\,J^{\textrm{\,c}}_{k}\,{\mathcal T}^{-1}\,=\,(-)^{k+1}\,J^{\textrm{\,c}}_{-k}\,$ and $\,{\mathcal T}\,Q^{\textrm{\,c}}_{k}\,{\mathcal T}^{-1}\,=\,(-)^{k}\,Q^{\textrm{\,c}}_{-k}\,$.  $Q^{\textrm{\,c}}_{k}$ can also be expressed in terms of the angular momentum operators, as on the right-hand side of Eq.\ \eqref{eq:Q} with $J^{\textrm{\,c}}_x\,\equiv\,\left(J^{\textrm{\,c}}_+\,+\,J^{\textrm{\,c}}_-\right)/2$ and $J^{\textrm{\,c}}_y\,\equiv\,\left(J^{\textrm{\,c}}_+\,-\,J^{\textrm{\,c}}_-\right)/2i$ (the bars over products of two
$J^{\textrm{\,c}}_k$s represent full symmetrization with respect to the different indices).

For reference, the octupolar electron–hole tensor operator is also constructed, as its components are needed for the spherical-symmetry-breaking terms in the next subsection.
\bea
\left(\begin{array}{c}
  O^{\textrm{\,c}}_3\\
  O^{\textrm{\,c}}_2\\
  O^{\textrm{\,c}}_1\\
  O^{\textrm{\,c}}_{0}\\
  O^{\textrm{\,c}}_{-1}\\
  O^{\textrm{\,c}}_{-2}\\
  O^{\textrm{\,c}}_{-3}
\end{array}\right)&\equiv&
\left[\begin{array}{c}
    \psi_{\frac 3 2}^{\,\Gamma_8\,\dagger\,}\psi_{-\frac 3 2}^{\,\Gamma_8}\\
    -\frac{ 1}{\sqrt 2}\left(\psi_{\frac 3 2}^{\,\Gamma_8\,\dagger\,}\psi_{-\frac 1 2}^{\,\Gamma_8}\,-\,\psi_{\frac 1 2}^{\,\Gamma_8\,\dagger\,}\psi_{-\frac 3 2}^{\,\Gamma_8}\right)\\
    \frac{ 1}{\sqrt 5}\left(\psi_{\frac 3 2}^{\,\Gamma_8\,\dagger\,}\psi_{\frac 1 2}^{\,\Gamma_8}\,-\,\sqrt{3}\,\psi_{\frac 1 2}^{\,\Gamma_8\,\dagger\,}\psi_{-\frac 1 2}^{\,\Gamma_8}\,+\,\psi_{-\frac 1 2}^{\,\Gamma_8\,\dagger\,}\psi_{-\frac 3 2}^{\,\Gamma_8}\right)\\
      -\frac{1}{2\sqrt{5}}\left(\psi_{\frac 3 2}^{\,\Gamma_8\,\dagger\,}\psi_{\frac 3 2}^{\,\Gamma_8}\,-\,3\,\psi_{\frac 1 2}^{\,\Gamma_8\,\dagger\,}\psi_{\frac 1 2}^{\,\Gamma_8}\,+\,3\,\psi_{-\frac 1 2}^{\,\Gamma_8\,\dagger\,}\psi_{-\frac 1 2}^{\,\Gamma_8}\,-\,\psi_{-\frac 3 2}^{\,\Gamma_8\,\dagger\,}\psi_{-\frac 3 2}^{\,\Gamma_8}\right)\\
      -\frac{ 1}{\sqrt 5}\left(\psi_{\frac 1 2}^{\,\Gamma_8\,\dagger\,}\psi_{\frac 3 2}^{\,\Gamma_8}\,-\,\sqrt{3}\,\psi_{-\frac 1 2}^{\,\Gamma_8\,\dagger\,}\psi_{\frac 1 2}^{\,\Gamma_8}\,+\,\psi_{-\frac 3 2}^{\,\Gamma_8\,\dagger\,}\psi_{-\frac 1 2}^{\,\Gamma_8}\right)\\
      -\frac{ 1}{\sqrt 2}\left(\psi_{-\frac 1 2}^{\,\Gamma_8\,\dagger\,}\psi_{\frac 3 2}^{\,\Gamma_8}\,-\,\psi_{-\frac 3 2}^{\,\Gamma_8\,\dagger\,}\psi_{\frac 1 2}^{\,\Gamma_8}\right)\\
      -\psi_{-\frac 3 2}^{\,\Gamma_8\,\dagger\,}\psi_{\frac 3 2}^{\,\Gamma_8}
  \end{array}\right].
\eea
Note that each octupolar component can also be written in terms of triple products of $J^{\textrm{\,c}}_k$s. These expressions are straightforward to derive and are omitted here as their explicit form is not used in the calculations.

\begin{figure}
\includegraphics[width=1\linewidth]{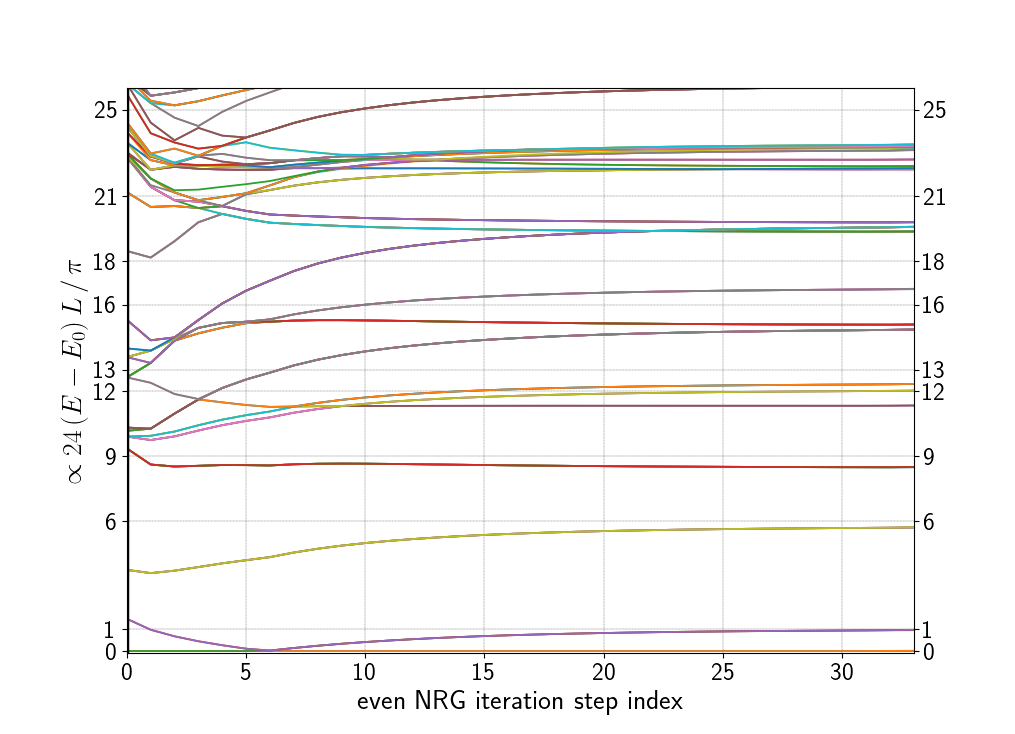}
\caption{Finite-size spectrum  of the spherically symmetric, quadrupolar exchange interaction, ${\cal H}^{\,\textrm{Q}}$, as the function of the NRG iteration step index, $N$, corresponding to a temperature, $T_N \approx \Lambda^{-N/2} D / k_B$, as extracted from NRG \cite{Toth08}, with $L$ the size of the chiral fermion system, $\Lambda=2$ chosen for the discretization parameter, $k_B$ the Boltzmann constant, and  $D$ the bandwidth \cite{Wilson75}.
 Only the U(1)$_\textrm{charge}$ symmetry of the model was exploited, and a minimum of 3000 multiplets were kept. A larger symmetry, SU(2)$_\textrm{spin}\times$U(1)$_\textrm{charge}$ could be used to obtain more accurate NRG results. The lowest levels still match the previous NRG results from Ref.\ \onlinecite{Koga99}.  The value of the dimensionless Kondo coupling is $\,D{\cal J}^{\textrm{Q}}=0.5\,$. The finite-size fixed point spectrum is independent of the value and sign of ${\cal J}^{\textrm{Q}}$ as long as ${\cal J}^{\textrm{Q}}\neq 0$.
}
\label{fig:qq}
\end{figure}

\begin{figure}
\includegraphics[width=1\linewidth]{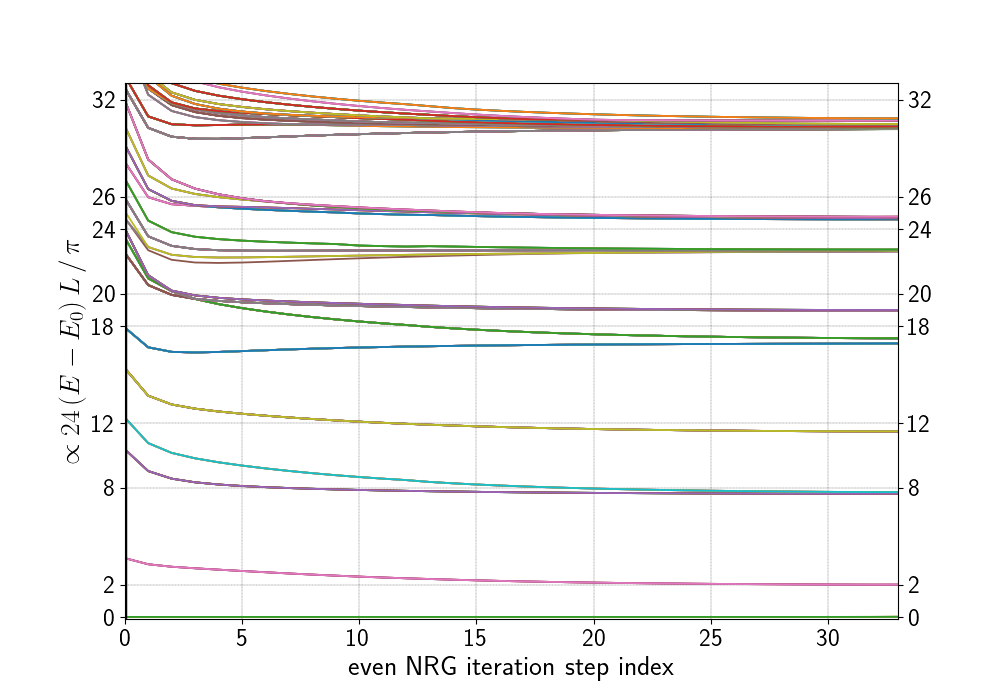}
\caption{Finite-size spectrum of the spherically symmetric, dipolar, spin exchange interaction, ${\cal H}^{\,\textrm{Kondo}}$,  as the function of the NRG iteration step index, $N$ with $\Lambda=2$ used for the discretization parameter \cite{Wilson75}.
  Only the U(1)$_\textrm{charge}$ symmetry of the model was exploited, and a minimum of 1500 multiplets were kept. A larger symmetry, SU(2)$_\textrm{spin}\times$U(1)$_\textrm{charge}$ could be used to get more accurate NRG results, but the agreement with the conformal field theory (CFT) results \cite{Koga99} for the lowest levels is still evident.  The numbers on the y-axes are the exact excitation energies from CFT.
  The value of the dimensionless Kondo coupling is $\,D{\cal J}^{\textrm{K}}=0.5\,$,  with $D$ the bandwidth. The finite-size fixed point spectrum is independent of the value of ${\cal J}^{\textrm{K}}$ as long as ${\cal J}^{\textrm{K}}>0$. For ${\cal J}^{\textrm{K}}<0$, the system is a Fermi liquid at low temperatures.   
}
\label{fig:spinspin}
\end{figure}

Thus, the following three independent terms can appear in a spherically symmetric exchange Hamiltonian, which is also cubic and time-reversal invariant, and describes the screening of a $J=1$ impurity by local $J=\frac 3 2$ conduction electrons
\bea
{\cal H}^{\,\textrm{PS}}&=&{\cal J}^{\textrm{PS}}\, n^{\textrm{\,imp}}\,n^{\textrm{\,c}}\\
{\cal H}^{\,\textrm{Kondo}}&=&{\cal J}^{\textrm{K}}\,\left[\frac 1 2\left(S^{\textrm{\,imp}}_+J^{\textrm{\,c}}_{-}\,+\,S^{\textrm{\,imp}}_-J^{\textrm{\,c}}_{+}\right)
  \,+\,S^{\textrm{\,imp}}_zJ^{\textrm{\,c}}_{z}\right]\\
{\cal H}^{\,\textrm{Q}}&=&{\cal J}^{\textrm{Q}}\,\sum_{k=-2}^2(-)^k\,Q^{\textrm{\,imp}}_k  Q^{\textrm{\,c}}_{-k}\,.
\eea
Solving
\bea
{\cal H}&=&\sum_{{\vec k},\,J_z}\epsilon({\vec k})\,c_{\,J=\frac 3 2\,J_z}^{\,\dagger}({\vec k})c_{\,J=\frac 3 2\,J_z}^{}({\vec k})\,+\,{\cal H}^{\,\textrm{PS}}\,+\,{\cal H}^{\,\textrm{Kondo}}\,+\,{\cal H}^{\,\textrm{Q}}\,,
\eea
in Ref.\ \onlinecite{Koga99}, by applying NRG and a perturbative scaling analysis, it was found that for  ${\cal J}^{\textrm{Q}}\neq 0$, the quadrupolar interaction dominates over the dipolar spin exchange and the model flows to an intermediate coupling, NFL fixed point. Its finite-size fixed-point spectrum (see Fig.\ \ref{fig:qq}) and the scaling dimension of its leading irrelevant operator, $\,\Delta\,=\,1/6\,$, were obtained using NRG. This scaling dimension implies that the impurity contribution to the specific heat coefficient and to the magnetic susceptibility both go as $T^{-2/3}$ for low $T$.  For ${\cal J}^{\textrm{K}}\neq 0, {\cal J}^{\textrm{Q}}=0$\,, the model flows to another intermediate coupling, NFL fixed point with a different level-spacing structure (see Fig.\ \ref{fig:spinspin}), and whose spin-sector can be mapped to that of the ten-channel Kondo model \cite{Fabrizio96,Sengupta96}. Nevertheless, using conformal field theory (CFT) \cite{Koga99}, it was found to have the same $\,\Delta\,=\,1/6\,$ scaling dimension for its leading irrelevant operator implying the same temperature dependence for the specific heat coefficient and the magnetic susceptibility. The potential scattering term, ${\cal H}^{\,\textrm{PS}}$, is exactly marginal around both NFL fixed points.  At the level of the finite-size spectrum, this operator appears through a phase shift $\delta$, i.e.\ as a shift of the energy levels.

\subsection{\label{subsec:CubSym} Exchange couplings with only cubic symmetry}
The construction of the cubic exchange couplings proceeds the same way as in Subsec.\ \ref{subsec:SpherSym}, with the difference that the spherical tensor product decompositions and the spherical Clebsch--Gordan coefficients are replaced by the cubic ones.  For a $\Gamma_4$ triplet impurity ground state, we can form four cubic, irreducible tensor operators from the ket and bra states according to the cubic decomposition rule \cite{Koster63}: $\Gamma_4\otimes\Gamma_4=\Gamma_1\oplus\Gamma_3\oplus\Gamma_4\oplus\Gamma_5$. We now make a rotation, and write the $\Gamma_4$ basis states as (c.f.\ Eq.\ \eqref{eq:spin1multiplet})
\bea
\left(
\begin{array}{l}
  |S_x\rangle\equiv\frac{-|1\,1\rangle+|1\,-1\rangle}{\sqrt2}\\
  |S_y\rangle\equiv\frac{-|1\,1\rangle-|1\,-1\rangle}{i\sqrt2}\\
  |S_z\rangle\equiv|1\quad\,0\rangle
\end{array}
\right)\,,
\quad
\left(
\begin{array}{l}
  \langle S_x|\equiv\frac{-\langle1\,1|+\langle1\,-1|}{\sqrt2}\\
  \langle S_y|\equiv\frac{\langle1\,1|+\langle1\,-1|}{i\sqrt2}\\
  \langle S_z|\equiv\langle1\quad\,0|
\end{array}
\right)\,,
\eea
so that we can make use of the cubic Clebsch--Gordan coefficients tabulated in Ref.\ \onlinecite{Koster63}. We find that the cubic scalar, i.e., the $\Gamma_1$ irreducible tensor operator, is the impurity number operator $n^{\textrm{imp}}$ defined in Eq.\ \eqref{eq:impno}, since invariance under continuous rotations implies invariance under the octahedral and tetrahedral groups. The two components of the $\Gamma_3$ irreducible tensor operator can be expressed as 
\begin{multline}
{\vec\Phi}^{\textrm{imp}\,\,3}\equiv
\left(\begin{array}{c}
  \Phi^{\textrm{imp}\,\,3}_1\\
  \Phi^{\textrm{imp}\,\,3}_2
\end{array}\right)\,\equiv\,
\left[\begin{array}{c}
  \frac{1}{\sqrt{6}}\left(|1\quad\,1\rangle \langle1\quad\,1|\,-\,2|1\quad\,0\rangle \langle1\quad\,0|\,+\,|1\,-1\rangle \langle1\,-1|\right)\\
 -\frac{1}{\sqrt{2}}\left(|1\quad\,1\rangle \langle1\,-1|\,+\,|1\,-1\rangle \langle1\quad\,1|\right)
  \end{array}
  \right]\\=
\left[\begin{array}{c}
  Q^{\textrm{\,imp}}_{0}\\
 - \frac{Q^{\textrm{\,imp}}_{2}\,+\,Q^{\textrm{\,imp}}_{-2}}{\sqrt{2}}
\end{array}
  \right]=
\left\{\begin{array}{c}
  \frac{1}{\sqrt{6}}\,\left[2\,\left(S^{\textrm{\,imp}}_z\right)^2\,-\,\left(S^{\textrm{\,imp}}_x\right)^2\,-\,\left(S^{\textrm{\,imp}}_y\right)^2\right]\\
  -\frac{1}{\sqrt{2}}\,\left[\left(S^{\textrm{\,imp}}_x\right)^2\,-\,\left(S^{\textrm{\,imp}}_y\right)^2\right]
\end{array}\right\}\,,
\end{multline}
and they are time-reversal even.
The three components of the $\Gamma_4$ and $\Gamma_5$ irreducible tensor operators are given by
\begin{multline}
    {\vec\Phi}^{\textrm{imp}\,\,4}\,\equiv\,
\left(\begin{array}{c}
  \Phi^{\textrm{imp}\,\,4}_x\\
  \Phi^{\textrm{imp}\,\,4}_y\\
  \Phi^{\textrm{imp}\,\,4}_z
\end{array}\right)\,\equiv\,
\left[\begin{array}{c}
    \frac{1}{2}\left(|1\quad\,1\rangle \langle1\quad\,0|\,+\,|1\quad\,0\rangle \langle1\,-1|\right)\,+\,\textrm{H.c.}\\
    -\frac{i}{2}\left(|1\quad\,1\rangle \langle1\quad\,0|\,+\,|1\quad\,0\rangle \langle1\,-1|\right)\,+\,\textrm{H.c.}\\
    \frac{1}{\sqrt{2}}\left(|1\quad\,1\rangle \langle1\quad\,1|\,-\,|1\,-1\rangle \langle1\,-1|\right)
  \end{array}
  \right]\,=\,\frac{{\vec S}^{\textrm{\,imp}}}{\sqrt{2}}\,,
\end{multline}
and
\begin{multline}
{\vec\Phi}^{\textrm{imp}\,\,5}\,\equiv\,
\left(\begin{array}{c}
  \Phi^{\textrm{imp}\,\,5}_{yz}\\
  \Phi^{\textrm{imp}\,\,5}_{zx}\\
  \Phi^{\textrm{imp}\,\,5}_{xy}
\end{array}\right)\,\equiv\,
\left[\begin{array}{c}
    -\frac{i}{2}\left(|1\quad\,1\rangle \langle1\quad\,0|\,-\,|1\quad\,0\rangle \langle1\,-1|\right)\,+\,\textrm{H.c.}\\
    \frac{1}{2}\left(|1\quad\,1\rangle \langle1\quad\,0|\,-\,|1\quad\,0\rangle \langle1\,-1|\right)\,+\,\textrm{H.c.}\\
    -\frac{i}{\sqrt{2}}\left(|1\quad\,1\rangle \langle1\,-1|\,-\,|1\,-1\rangle \langle1\quad\,1|\right)
  \end{array}
  \right]\\=\,\left(\begin{array}{c}
    \frac{Q^{\textrm{\,imp}}_{1}\,+\,Q^{\textrm{\,imp}}_{-1}}{-i\,\sqrt{2}}\\
    \frac{-Q^{\textrm{\,imp}}_{1}\,+\,Q^{\textrm{\,imp}}_{-1}}{\sqrt{2}}\\
  \frac{Q^{\textrm{\,imp}}_{2}\,-\,Q^{\textrm{\,imp}}_{-2}}{i\,\sqrt{2}}
\end{array}
\right)\,=-\frac{1}{\sqrt 2}\left(\begin{array}{c}
    \overline{S^{\textrm{\,imp}}_y\,S^{\textrm{\,imp}}_z}\\
    \overline{S^{\textrm{\,imp}}_x\,S^{\textrm{\,imp}}_z}\\
    \overline{S^{\textrm{\,imp}}_x\,S^{\textrm{\,imp}}_y}
\end{array}\right)
\,,
\end{multline}
so $\,{\vec\Phi}^{\textrm{\,imp}\,\,4}\,$ is time-reversal odd, whereas
  ${\vec\Phi}^{\textrm{\,imp}\,\,5}$ is time-reversal invariant. 

As the next step, we construct the cubic, conduction electron-hole, irreducible tensor operators from the tensor product of ${\bm \Psi^{\Gamma_8\,\dagger}}$ and
${\bm \Psi^{\Gamma_8}}$ given in Eq.\ \eqref{eq:conde_quartet} using the formula $\Gamma_8\otimes\Gamma_8\,=\,\Gamma_1\oplus\Gamma_2\oplus\Gamma_3\oplus2\Gamma_4\oplus\Gamma_5$. Except for the $\Gamma_2$ term,  all the terms on the right-hand side can be used to build cubic scalar exchange Hamiltonians. The $\Gamma_1$ term is the conduction electron number operator, $n^{\textrm{c}}$, given by Eq.\ \eqref{eq:condeno}. The two components of the $\Gamma_3$, conduction electron-hole irreducible tensor operator are
\begin{multline}
{\vec\Phi}^{\textrm{c}\,\,3}\equiv
\left(\begin{array}{c}
  \Phi^{\textrm{c}\,\,3}_1\\
  \Phi^{\textrm{c}\,\,3}_2
\end{array}\right)\equiv
\frac{1}{2}\left(\begin{array}{c}
  \psi_{\frac 3 2}^{\,\Gamma_8\,\dagger\,}\psi_{\frac 3 2}^{\,\Gamma_8}\,-\,\psi_{\frac 1 2}^{\,\Gamma_8\,\dagger\,}\psi_{\frac 1 2}^{\,\Gamma_8}\,-\,\psi_{-\frac 1 2}^{\,\Gamma_8\,\dagger\,}\psi_{-\frac 1 2}^{\,\Gamma_8}\,+\,\psi_{-\frac 3 2}^{\,\Gamma_8\,\dagger\,}\psi_{-\frac 3 2}^{\,\Gamma_8}\\
\psi_{\frac 3 2}^{\,\Gamma_8\,\dagger\,}\psi_{-\frac 1 2}^{\,\Gamma_8}\,+\,\psi_{\frac 1 2}^{\,\Gamma_8\,\dagger\,}\psi_{-\frac 3 2}^{\,\Gamma_8}\,+\,\psi_{-\frac 1 2}^{\,\Gamma_8\,\dagger\,}\psi_{\frac 3 2}^{\,\Gamma_8}\,+\,\psi_{-\frac 3 2}^{\,\Gamma_8\,\dagger\,}\psi_{\frac 1 2}^{\,\Gamma_8}
  \end{array}
\right)\\=\,
\left(\begin{array}{c}
  Q^{\textrm{\,c}}_0\\
  -\frac{Q^{\textrm{\,c}}_2\,+\,Q^{\textrm{\,c}}_{-2}}{\sqrt{2}}
\end{array}\right)\,=\,
\left[\begin{array}{c}
  \frac{1}{\sqrt{6}}\,\left(2\,\left(J^{\textrm{\,c}}_z\right)^2\,-\,\left(J^{\textrm{\,c}}_x\right)^2\,-\,\left(J^{\textrm{\,c}}_y\right)^2\right)\\
  -\frac{1}{\sqrt{2}}\,\left[\left(J^{\textrm{\,c}}_x\right)^2\,-\,\left(J^{\textrm{\,c}}_y\right)^2\right]
\end{array}\right]\,.
  \end{multline}
They are invariant under time-reversal.
We have the following two $\Gamma_4$ irreducible tensor operators
\begin{multline}
 {\vec\Phi}^{c\,\,4}\equiv
\left(\begin{array}{c}
  \Phi^{\,\textrm{c}\,\,4}_x\\
  \Phi^{\,\textrm{c}\,\,4}_y\\
  \Phi^{\,\textrm{c}\,\,4}_z
\end{array}\right)\,\equiv\,\left[\begin{array}{c}
    \frac{\sqrt{3}}{2}\,\psi_{\frac 3 2}^{\,\Gamma_8\,\dagger\,}\psi_{\frac 1 2}^{\,\Gamma_8}\,+\,\psi_{\frac 1 2}^{\,\Gamma_8\,\dagger\,}\psi_{-\frac 1 2}^{\,\Gamma_8}\,+\,\frac{\sqrt{3}}{2}\,\psi_{-\frac 1 2}^{\,\Gamma_8\,\dagger\,}\psi_{-\frac 3 2}^{\,\Gamma_8}\,+\,\textrm{H.c.}\\
    -i\left(\frac{\sqrt{3}}{2}\,\psi_{\frac 3 2}^{\,\Gamma_8\,\dagger\,}\psi_{\frac 1 2}^{\,\Gamma_8}\,+\,\psi_{\frac 1 2}^{\,\Gamma_8\,\dagger\,}\psi_{-\frac 1 2}^{\,\Gamma_8}\,+\,\frac{\sqrt{3}}{2}\,\psi_{-\frac 1 2}^{\,\Gamma_8\,\dagger\,}\psi_{-\frac 3 2}^{\,\Gamma_8}\right)\,+\,\textrm{H.c.}\\
    \frac{3}{2}\,\psi_{\frac 3 2}^{\,\Gamma_8\,\dagger\,}\psi_{\frac 3 2}^{\,\Gamma_8}\,+\,\frac{1}{2}\,\psi_{\frac 1 2}^{\,\Gamma_8\,\dagger\,}\psi_{\frac 1 2}^{\,\Gamma_8}\,-\,\frac{1}{2}\,\psi_{-\frac 1 2}^{\,\Gamma_8\,\dagger\,}\psi_{-\frac 1 2}^{\,\Gamma_8}\,-\,\frac{3}{2}\,\psi_{-\frac 3 2}^{\,\Gamma_8\,\dagger\,}\psi_{-\frac 3 2}^{\,\Gamma_8}
  \end{array}
  \right]\\=\,\left[\begin{array}{c}
  J^{\textrm{\,c}}_x\equiv \left(J^{\textrm{\,c}}_+\,+\,J^{\textrm{\,c}}_-\right)/2\\
  J^{\textrm{\,c}}_{y}\equiv \left(J^{\textrm{\,c}}_+\,-\,J^{\textrm{\,c}}_-\right)/2i\\
  J^{\textrm{\,c}}_z
\end{array}\right]\,,
\end{multline}
and
\begin{multline}
{\vec\Upsilon}^{\,\textrm{c}\,\,4}\equiv
\left(\begin{array}{c}
  \Upsilon^{\,\textrm{c}\,\,4}_x\\
  \Upsilon^{\,\textrm{c}\,\,4}_y\\
  \Upsilon^{\,\textrm{c}\,\,4}_z
\end{array}\right)\propto
\left[\begin{array}{c}
    \frac{1}{4}\left(\sqrt{5}\,\psi_{\frac 3 2}^{\,\Gamma_8\,\dagger\,}\psi_{-\frac 3 2}^{\,\Gamma_8}\,-\,\frac{\sqrt{3}}{\sqrt{5}}\,\psi_{\frac 3 2}^{\,\Gamma_8\,\dagger\,}\psi_{\frac 1 2}^{\,\Gamma_8}\,+\,\frac{3}{\sqrt{5}}\,\psi_{\frac 1 2}^{\,\Gamma_8\,\dagger\,}\psi_{-\frac 1 2}^{\,\Gamma_8}\,-\,\frac{\sqrt{3}}{\sqrt{5}}\,\psi_{-\frac 1 2}^{\,\Gamma_8\,\dagger\,}\psi_{-\frac 3 2}^{\,\Gamma_8}\right)\,+\,\textrm{H.c.}\\
    \frac{i}{4}\left(\sqrt{5}\,\psi_{\frac 3 2}^{\,\Gamma_8\,\dagger\,}\psi_{-\frac 3 2}^{\,\Gamma_8}\,+\,\frac{\sqrt{3}}{\sqrt{5}}\,\psi_{\frac 3 2}^{\,\Gamma_8\,\dagger\,}\psi_{\frac 1 2}^{\,\Gamma_8}\,-\,\frac{3}{\sqrt{5}}\,\psi_{\frac 1 2}^{\,\Gamma_8\,\dagger\,}\psi_{-\frac 1 2}^{\,\Gamma_8}\,+\,\frac{\sqrt{3}}{\sqrt{5}}\,\psi_{-\frac 1 2}^{\,\Gamma_8\,\dagger\,}\psi_{-\frac 3 2}^{\,\Gamma_8}\right)\,+\,\textrm{H.c.}\\
    \frac{1}{2\sqrt{5}}\left(\psi_{\frac 3 2}^{\,\Gamma_8\,\dagger\,}\psi_{\frac 3 2}^{\,\Gamma_8}\,-\,3\,\psi_{\frac 1 2}^{\,\Gamma_8\,\dagger\,}\psi_{\frac 1 2}^{\,\Gamma_8}\,+\,3\,\psi_{-\frac 1 2}^{\,\Gamma_8\,\dagger\,}\psi_{-\frac 1 2}^{\,\Gamma_8}\,-\,\psi_{-\frac 3 2}^{\,\Gamma_8\,\dagger\,}\psi_{-\frac 3 2}^{\,\Gamma_8}\right)
  \end{array}
  \right]\\=
\left[\begin{array}{c}
    \frac{1}{4}\left(\sqrt{5}\,O^{\textrm{\,c}}_3\,-\,\sqrt{3}\,O^{\textrm{\,c}}_1\right)\,+\,\textrm{H.c.}\\
    \frac{i}{4}\left(\sqrt{5}\,O^{\textrm{\,c}}_3\,+\,\sqrt{3}\,O^{\textrm{\,c}}_1\right)\,+\,\textrm{H.c.}\\
    O^{\textrm{\,c}}_0
\end{array}\right]
\,.
\end{multline}
They are both time-reversal odd, but the components of ${\vec\Upsilon}^{c\,\,4}$ are not linear combinations of the components of ${\vec J}^{\,c}$ or ${\vec Q}^{\,c}$\, as they contain the octupolar term $\psi_{\frac 3 2}^{\,\Gamma_8\,\dagger\,}\psi_{-\frac 3 2}^{\,\Gamma_8}$, for example. They are linear combinations of the octupolar electron-hole tensor operator. 
The two $\Gamma_5$ tensors can be defined as 
\begin{multline}
{\vec\Phi}^{\,\textrm{c}\,\,5}\equiv
\left(\begin{array}{c}
  \Phi^{\,\textrm{c}\,\,5}_{yz}\\
  \Phi^{\,\textrm{c}\,\,5}_{zx}\\
  \Phi^{\,\textrm{c}\,\,5}_{xy}
\end{array}\right)\propto
\left[\begin{array}{c}
    \frac{i}{2}\left(-\psi_{\frac 3 2}^{\,\Gamma_8\,\dagger\,}\psi_{\frac 1 2}^{\,\Gamma_8}\,+\,\psi_{-\frac 1 2}^{\,\Gamma_8\,\dagger\,}\psi_{-\frac 3 2}^{\,\Gamma_8}\right)\,
    +\,\textrm{H.c.}\\
    \frac{1}{2}\left(\psi_{\frac 3 2}^{\,\Gamma_8\,\dagger\,}\psi_{\frac 1 2}^{\,\Gamma_8}\,-\,\psi_{-\frac 1 2}^{\,\Gamma_8\,\dagger\,}\psi_{-\frac 3 2}^{\,\Gamma_8}\right)\,
    +\,\textrm{H.c.}\\
    -\frac{i}{2}\left(\psi_{\frac 3 2}^{\,\Gamma_8\,\dagger\,}\psi_{-\frac 1 2}^{\,\Gamma_8}\,+\,\psi_{\frac 1 2}^{\,\Gamma_8\,\dagger\,}\psi_{-\frac 3 2}^{\,\Gamma_8}\right)\,
    +\,\textrm{H.c.}
  \end{array}
  \right]\\=\frac{1}{\sqrt 2}\left[\begin{array}{c}
    -\,i\,\left(Q^{\,\textrm{c}}_1\,+\,Q^{\,\textrm{c}}_{-1}\right)\\
    Q^{\,\textrm{c}}_1\,-\,Q^{\,\textrm{c}}_{-1}\\
    i\,\left(Q^{\,\textrm{c}}_2\,-\,Q^{\,\textrm{c}}_{-2}\right)
\end{array}\right]=-\frac{1}{\sqrt 2}\left(\begin{array}{c}
    \overline{J^{\textrm{\,c}}_y\,J^{\textrm{\,c}}_z}\\
    \overline{J^{\textrm{\,c}}_x\,J^{\textrm{\,c}}_z}\\
    \overline{J^{\textrm{\,c}}_x\,J^{\textrm{\,c}}_y}
\end{array}\right)\,,
\end{multline}
which is time-reversal even, and
\begin{multline}
{\vec\Upsilon}^{\,\textrm{c}\,\,5}\equiv
\left(\begin{array}{c}
  \Upsilon^{\,\textrm{c}\,\,5}_{yz}\\
  \Upsilon^{\,\textrm{c}\,\,5}_{zx}\\
  \Upsilon^{\,\textrm{c}\,\,5}_{xy}
\end{array}\right)\propto
\left[\begin{array}{c}
    \frac{1}{4}\left(\sqrt{3}\,\psi_{\frac 3 2}^{\,\Gamma_8\,\dagger\,}\psi_{-\frac 3 2}^{\,\Gamma_8}\,+\,\psi_{\frac 3 2}^{\,\Gamma_8\,\dagger\,}\psi_{\frac 1 2}^{\,\Gamma_8}\,-\,\sqrt{3}\,\psi_{\frac 1 2}^{\,\Gamma_8\,\dagger\,}\psi_{-\frac 1 2}^{\,\Gamma_8}\,+\,\psi_{-\frac 1 2}^{\,\Gamma_8\,\dagger\,}\psi_{-\frac 3 2}^{\,\Gamma_8}\right)\,+\,\textrm{H.c.}\\
    -\frac{i}{4}\left(\sqrt{3}\,\psi_{\frac 3 2}^{\,\Gamma_8\,\dagger\,}\psi_{-\frac 3 2}^{\,\Gamma_8}\,-\,\psi_{\frac 3 2}^{\,\Gamma_8\,\dagger\,}\psi_{\frac 1 2}^{\,\Gamma_8}\,+\,\sqrt{3}\,\psi_{\frac 1 2}^{\,\Gamma_8\,\dagger\,}\psi_{-\frac 1 2}^{\,\Gamma_8}\,-\,\psi_{-\frac 1 2}^{\,\Gamma_8\,\dagger\,}\psi_{-\frac 3 2}^{\,\Gamma_8}\right)\,+\,\textrm{H.c.}\\
    \frac{1}{2}\left(-\,\psi_{\frac 3 2}^{\,\Gamma_8\,\dagger\,}\psi_{-\frac 1 2}^{\,\Gamma_8}\,+\,\psi_{\frac 1 2}^{\,\Gamma_8\,\dagger\,}\psi_{-\frac 3 2}^{\,\Gamma_8}\right)\,+\,\textrm{H.c.}
  \end{array}
  \right]\\
=\left[\begin{array}{c}
    \frac 1 4 \left(\sqrt{3}\,O^{\textrm{\,c}}_3\,+\,\sqrt{5}\,O^{\textrm{\,c}}_1\right)\,+\,\textrm{H.c.}\\
    -\frac i 4 \left(\sqrt{3}\,O^{\textrm{\,c}}_3\,-\,\sqrt{5}\,O^{\textrm{\,c}}_1\right)\,+\,\textrm{H.c.}\\
    \frac{1}{\sqrt{2}}\,O^{\textrm{\,c}}_2\,+\,\textrm{H.c.}
\end{array}\right]\,,
\end{multline}
which is time-reversal odd.
Thus, we have the following six independent, cubic-invariant exchange couplings
\bea
    {\cal H}^{\,\textrm{PS}}_{\,\Gamma_1\otimes\Gamma_1}&=&{\cal J}^{\,\textrm{PS}}_{\,\Gamma_1\otimes\Gamma_1}\, n^{\textrm{\,imp}}\,n^{\textrm{\,c}}\\
     {\cal H}^{\,\textrm{Q-Q}}_{\,\Gamma_3\otimes\Gamma_3}&=&{\cal J}^{\,\textrm{Q-Q}}_{\,\Gamma_3\otimes\Gamma_3}\,\,\,{\vec\Phi}^{\textrm{\,imp}\,\,3}\cdot\,{\vec\Phi}^{\textrm{\,c}\,\,3\,}_i\\
      {\cal H}^{\,\textrm{Kondo}}_{\,\Gamma_4\otimes\Gamma_4}&=&{\cal J}^{\,\textrm{K}}_{\,\Gamma_4\otimes\Gamma_4}\,\,\,{\vec \Phi}^{\textrm{\,imp}\,\,4\,}\cdot\,{\vec \Phi}^{\textrm{\,c}\,\,4}\\
     {\cal H}^{\,\textrm{D-O}}_{\,\Gamma_4\otimes\Gamma_4}&=&{\cal J}^{\,\textrm{D-O}}_{\,\Gamma_4\otimes\Gamma_4}\,\,\,{\vec \Phi}^{\textrm{\,imp}\,\,4\,}\cdot\,{\vec\Upsilon}^{\textrm{\,c}\,\,4}\\
     {\cal H}^{\,\textrm{Q-Q}}_{\,\Gamma_5\otimes\Gamma_5}&=&{\cal J}^{\,\textrm{Q-Q}}_{\,\Gamma_5\otimes\Gamma_5}\,\,\,{\vec\Phi}^{\textrm{\,imp}\,\,5\,}\cdot\,{\vec\Phi}^{\textrm{\,c}\,\,5}\\
     {\cal H}^{\,\textrm{Q-O}}_{\,\Gamma_5\otimes\Gamma_5}&=&{\cal J}^{\,\textrm{Q-O}}_{\,\Gamma_5\otimes\Gamma_5}\,\,\,{\vec\Phi}^{\textrm{\,imp}\,\,5\,}\cdot\,{\vec\Upsilon}^{\textrm{\,c}\,\,5}\,.
     \eea
     All of them are time-reversal invariant except for ${\cal H}^{\,\textrm{Q-O}}_{\,\Gamma_5\otimes\Gamma_5}$, which changes sign under time-reversal, similarly to Cox's 2CK Hamiltonian \cite{Toth25}. Whether this renders ${\cal H}^{\,\textrm{Q-O}}_{\,\Gamma_5\otimes\Gamma_5}$ or Cox's 2CK Hamiltonian unphysical is unclear, as, at low temperatures, in the Kondo regime, time-reversal invariance can be lost by losing access to the first excited states as in Cox's 2CK case, even though the underlying Anderson Hamiltonian is time-reversal invariant \cite{Cox87,Cox98}.

     ${\cal H}^{\,\textrm{PS}}_{\,\Gamma_1\otimes\Gamma_1}$ is the potential scattering, identical to ${\cal H}^{\,\textrm{PS}}$. It is an exactly marginal perturbation around all the NFL fixed points studied in this work.
The interaction $ {\cal H}^{\,\textrm{Q-Q}}_{\,\Gamma_3\otimes\Gamma_3}$ is quadrupolar in both the impurity and the conduction electron degrees of freedom. It is also marginal and appears  as a shift of the energy levels in the excitation spectrum. The magnitude of this shift depends on the value of  ${\cal J}^{\,\textrm{Q-Q}}_{\,\Gamma_3\otimes\Gamma_3}$. The interactions ${\cal H}^{\,\textrm{Kondo}}_{\,\Gamma_4\otimes\Gamma_4}$ and ${\cal H}^{\,\textrm{Kondo}}$ are identical apart from a constant factor of $\sqrt{2}$, and flow to the intermediate coupling, NFL fixed point with the level-spacing structure shown in Fig.\ \ref{fig:spinspin}. Even though the dipolar-octupolar exchange, ${\cal H}^{\,\textrm{D-O}}_{\,\Gamma_4\otimes\Gamma_4}$, breaks spherical symmetry in the high-temperature, local moment regime, it also flows to the same NFL fixed point as ${\cal H}^{\,\textrm{Kondo}}_{\,\Gamma_4\otimes\Gamma_4}$ for ${\cal J}^{\,\textrm{D-O}}_{\,\Gamma_4\otimes\Gamma_4}>0$, whereas for ${\cal J}^{\,\textrm{D-O}}_{\,\Gamma_4\otimes\Gamma_4}<0$, it is a Fermi liquid at low temperatures as confirmed using NRG \cite{Toth08}.

\begin{figure}
\includegraphics[width=1\linewidth]{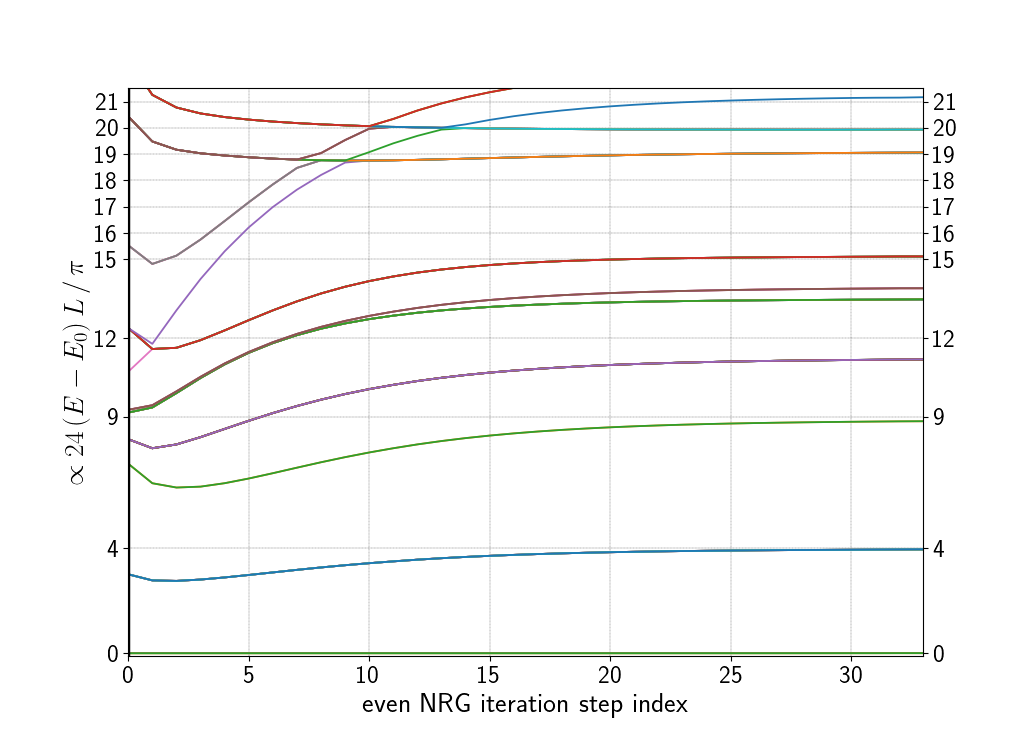}
\caption{Finite-size spectrum of the spherical symmetry breaking, quadrupolar ${\cal H}^{\,\Gamma_5\otimes\Gamma_5}_{\textrm{I}}$  as the function of the NRG iteration step index, $N$.  $\Lambda=2$ was chosen for the discretization parameter \cite{Wilson75}.
  The U(1)$_\textrm{charge}$ symmetry of the model was used, and a minimum of 5000 multiplets were kept.  The value of the dimensionless Kondo coupling is $\,D{\cal J}^{\,\textrm{Q-Q}}_{\,\Gamma_5\otimes\Gamma_5}=0.5\,$, with $D$ the bandwidth. However, the finite-size fixed point spectrum is independent of the value and sign of ${\cal J}^{\,\textrm{Q-Q}}_{\,\Gamma_5\otimes\Gamma_5}$ for ${\cal J}^{\,\textrm{Q-Q}}_{\,\Gamma_5\otimes\Gamma_5}\neq 0$.
}
\label{fig:G5}
\end{figure}

${\cal H}^{\,\textrm{Q-Q}}_{\,\Gamma_5\otimes\Gamma_5}$ is a time-reversal invariant, quadrupolar-quadrupolar exchange, whereas ${\cal H}^{\,\textrm{Q-O}}_{\,\Gamma_5\otimes\Gamma_5}$ is a time-reversal symmetry breaking, quadrupolar-octupolar exchange.
In our chosen basis, ${\cal H}^{\,\textrm{Q-O}}_{\,\Gamma_5\otimes\Gamma_5}$ is pure imaginary.  
However, while for ${\cal J}^{\,\textrm{Q-Q}}_{\,\Gamma_5\otimes\Gamma_5}\neq0$ and ${\cal J}^{\,\textrm{Q-O}}_{\,\Gamma_5\otimes\Gamma_5}\neq 0$, the sum of these interactions flows to the fixed point of the spherically symmetric quadrupolar exchange interaction, whereas for ${\cal J}^{\,\textrm{Q-Q}}_{\,\Gamma_5\otimes\Gamma_5}\neq0$ and ${\cal J}^{\,\textrm{Q-O}}_{\,\Gamma_5\otimes\Gamma_5}=0$, a novel NFL fixed point is found whose finite-size spectrum is shown in Fig.\ \ref{fig:G5}. This level structure is independent of the value of ${\cal J}^{\,\textrm{Q-Q}}_{\,\Gamma_5\otimes\Gamma_5}$. Around the fixed point of the spherically symmetric, quadrupolar exchange Hamiltonian, ${\cal H}^{\,\textrm{Q-Q}}_{\,\Gamma_5\otimes\Gamma_5}$ is irrelevant, but on its own, it flows to a previously unidentified NFL fixed point, at least in the context of triplet impurities.

  Several important questions remain open regarding this coupling, like  what the spherically symmetric model is that has the same NFL fixed point, and, assuming that the impurity contribution to the zero-point entropy is fractional, as it is for all known NFL exchange models, what kind of Kondo anyon emerges in its low-temperature phase. This latter question applies to ${\cal H}^{\,\textrm{Q}}$ too.  NRG could be used to extract the scaling dimension of the leading irrelevant operator, as it has been done for ${\cal H}^{\,\textrm{Q}}$. The scaling dimension can then be used to predict the temperature dependence of physical quantities, such as the impruty contribution to the specific heat, or various dynamical quantities.  Both models lack a conformal field theory solution as yet.

For the physical realizations of the corresponding NFLs in cubic metals, an impurity ion with a triplet ground state (i.e.\ with an even electron configuration) is required, then, provided that ${\cal J}^{\,\textrm{Q-Q}}_{\,\Gamma_5\otimes\Gamma_5}$ dominates over the other couplings, in a certain temperature range, the novel NFL behavior could be observed. Alternative realizations of the triplet-impurity NFL physics in quantum dot devices and ultracold atomic gases could also be explored.

\section{\label{sec:end}Conclusions}

With the aim of identifying all types of NFL quantum critical behavior that might appear in a cubic metal due to quantum impurities,   
I studied the cubic-symmetry-allowed exchange interactions between a magnetic moment with a triplet ground state and $\Gamma_8$ conduction electrons to complement our previous catalog of cubic symmetry-protected NFLs for doublet impurities \cite{Toth25}.  For a triplet impurity, previous studies were limited to the three spherically-symmetric exchange interactions, while here  all six, cubic-symmetry-allowed
couplings are constructed. I solved these interactions using NRG and found that with the appropriate sign for each coupling constant, each exchange interaction either leads to a NFL or is marginal, as is also the case for a doublet impurity in the same environment \cite{Toth25}.  One of the couplings, a quadrupolar-quadrupolar coupling which breaks spherical symmetry in the high-temperature local moment regime, has a novel NFL fixed point. This represents the third distinct NFL universality class available for triplet impurities in a cubic metal, complementing the two known cases.
This coupling is an irrelevant perturbation around the already known fixed point of the quadrupolar coupling.  Nevertheless it would be interesting to identify its spherically symmetric fixed point Hamiltonian and the Kondo anyon that emerges in its NFL phase, to understand whether it can mapped to a conventional multichannel Kondo model, as well as to explore its thermodynamic and dynamical properties.


\begin{acknowledgments}
This project has received funding from the
European Union’s Horizon 2020 research and innovation programme under the Marie Skłodowska-Curie Grant Agreement No.\ 101024548. I am grateful to Andrew Huxley and Andrew Mitchell for many useful discussions. 

\end{acknowledgments}

\section*{Data Availability Statement}

The data that support the findings of this study are available from the corresponding author upon reasonable request.

\nocite{*}

\end{document}